\documentclass[]{aastex631}

\usepackage{amsmath}  
\usepackage{tabularx} 
\usepackage{xcolor} 
\usepackage[colorinlistoftodos]{todonotes}
\usepackage{longtable}
\usepackage{threeparttable}

\begin{document}

\title{The Ongoing Decline in Activity of Comet 103P/Hartley 2}

\author[0000-0001-9003-0999]{Ariel Graykowski}
\affiliation{SETI Institute, 339 Bernardo Ave, Suite 200, Mountain View, CA 94043, USA}

\author[0009-0008-0066-4188]{Guillaume Langin}
\affiliation{Association Française d'Astronomie, 17 Rue Émile Deutsch de la Meurthe, 75014 Paris, France}
\affiliation{Ciel \& Espace, Paris, France}

\author[0000-0001-8100-1504]{David Chiron}
\affiliation{Association Française d’Astronomie Citizen Astronomer, France}

\author[0000-0003-4091-0247]{Bruno Guillet}
\affiliation{Unistellar Citizen Astronomer}

\author[0000-0001-7016-7277]{Franck Marchis}
\affiliation{SETI Institute, 339 Bernardo Ave, Suite 200, Mountain View, CA 94043, USA}
\affiliation{Unistellar, 5 All. Marcel Leclerc bâtiment C, 13008 Marseille, France}

\author[0000-0003-2414-5370]{Nicolas Biver}
\affiliation{Observatoire de Paris, 77 Av. Denfert Rochereau, 75014 Paris, France}

\author{Gérard Arlic}
\affiliation{Association Française d’Astronomie Citizen Astronomer, France}

\author{Bernard Baudouin}
\affiliation{Association Française d’Astronomie Citizen Astronomer, France}

\author{Etienne Bertrand}
\affiliation{Association Française d’Astronomie Citizen Astronomer, France}

\author{Randall Blake}
\affiliation{Unistellar Citizen Astronomer}

% ...

\begin{abstract}
We report photometric observations of Comet 103P/Hartley 2 during its 2023 apparition. Our campaign, conducted from August through December 2023, combined data from a global network of citizen astronomers coordinated by Unistellar and the Association Française d’Astronomie. Photometry was derived using an automated pipeline for eVscope observations in partnership with the SETI Institute and aperture photometry via AstroLab Stellar. We find that the comet’s peak reduced brightness, measured at $G_{\rm min} = 10.24 \pm 0.47$, continues a long-term fading trend since 1991. The decline in activity follows a per-apparition minimum magnitude increase of $\Delta G_{\rm min} = 0.59 \pm 0.11$ mag, corresponding to a $\sim$42\% reduction in brightness each return. This trend implies that the comet’s active fraction has declined by approximately an order of magnitude since 1991 and may indicate that Hartley 2 is no longer hyperactive by definition. The fading is consistent with progressive volatile depletion rather than orbital effects. These results offer insight into the evolutionary processes shaping Jupiter-family comets.
\end{abstract}


\section{Introduction} \label{sec:intro}

Comet 103P/Hartley 2 is a historically hyperactive Jupiter-family comet (JFC) with a short orbital period of approximately 6.5 years. Its frequent apparitions and high levels of activity have made it a key case study for understanding the long-term evolution of JFCs. Discovered in 1986 by Malcolm Hartley \citep{hartley1986}, the comet was visited by NASA’s EPOXI spacecraft during its 2010 apparition \citep{AHearn2011}, and was also the focus of an unprecedented international ground-based observing campaign during that apparition \citep{Meech2011}. Together, these observations have made Hartley 2 one of the most thoroughly studied comets in the Solar System.

The 2010 observations and spacecraft visit captured details of the comet’s complex, non-principal axis rotation \citep{samarasinha2011, knight2011, belton2013}, constrained its nucleus properties \citep{AHearn2011, Thomas2013}, and revealed compositional gradients and spatial heterogeneity in gas emissions \citep{DelloRusso2011, Mumma2011}. They also documented changes in the comet’s rotational period \citep{knight2015}, a phenomenon observed again during the 2023 apparition \citep{Lehmann2025}. Hartley 2’s nucleus is relatively small, with a bilobate shape, a long axis of 2.33 km, a minimum diameter of 0.69 km, and a mean radius of 0.58 km \citep{AHearn2011, Thomas2013}. Despite its size, the comet has shown unusually high levels of activity. This hyperactivity refers to the gas production rate exceeding what the nucleus's surface area alone could sustain, implying the presence of additional sources of volatiles. EPOXI revealed that this hyperactivity is driven by centimeter-sized icy chunks released from the smaller lobe likely entrained by $CO_2$-driven jets, which sublimate in the coma \citep{AHearn2011, Kelley2013}.

Over successive perihelion passages, Hartley 2 has shown a measurable decline in activity, likely due to the progressive depletion of accessible volatiles \citep{combi2011, knight2013}. This trend has been observed through photometric and spectroscopic studies, revealing reduced brightness and gas production rates between the 1991, 1997/98, and 2010/11 apparitions. The 2004 and 2017 apparitions were poorly placed for Earth-based observations, while 2023 provided better observing conditions (though still not as well-placed as 2010). In this work, we present new observations from the 2023 apparition and confirm that the comet’s long-term decline in activity is ongoing. Possible mechanisms for activity decline include volatile depletion, uneven distribution of volatiles, crust formation, and fragmentation or disintegration, each of which is discussed in detail in Section~\ref{Long-Term Fate of 103P/Hartley 2}.

Hartley 2’s fading activity can be contextualized by comparing it to other well-studied JFCs. One notable example is Comet 9P/Tempel 1 \citep{hacken1995a}, the target of NASA’s \textit{Deep Impact} mission \citep{ahearn2005}. Tempel 1’s declining activity has been attributed to rotational and solar insolation effects. Its low obliquity and large polar active region lead to uneven solar heating and asymmetric volatile depletion, while precessional motion gradually reduces the fraction of the surface receiving optimal solar input during successive apparitions \citep{schleicher2007}. Like Tempel 1, Hartley 2 exhibits pre-/post-perihelion asymmetry in brightness, with faster brightening before perihelion and more gradual fading after. However, unlike Tempel 1, Hartley 2’s complex, non-principal axis rotation makes it unlikely that solar illumination follows a consistent pattern from one apparition to the next \citep{samarasinha2011, knight2011, belton2013, knight2015}.

Understanding the long-term evolution of Hartley 2 is essential for characterizing the aging process of JFCs and provides a comparative framework for assessing other short-period comets. The declining activity seen in Hartley 2 resembles that of other comets such as 2P/Encke, 46P/Wirtanen, and 41P/Tuttle-Giacobini-Kresák, where volatile depletion appears to drive a secular decrease in activity over time \citep{haken1995, combi2020, knight2021}. Weakly active comets such as 209P/LINEAR, 252P/LINEAR, and P/2021 HS (PANSTARRS) are thought to represent more evolved stages of this process, where long-term aging has led to the near-exhaustion of accessible volatiles \citep{ye2016a, ye2016b, ye2023}. Future studies incorporating high-resolution photometric and spectroscopic observations will be essential for probing the physical and compositional evolution driving the activity decline in 103P and other short-period comets.

\section{Observations} \label{sec:obs}

During its 2023 apparition, we leveraged global networks of citizen astronomers to monitor the activity of Hartley 2. The Unistellar eVscope has been widely adopted as a tool for citizen science observations, enabling real-time data collection by amateur astronomers worldwide \citep{marchis2020}. Additionally, the Association Française d’Astronomie (AFA) has played a key role in collaborative citizen science projects, engaging amateur astronomers in coordinated research efforts \citep{heutte2017}. For this study, we utilized the Unistellar Network in conjunction with participating AFA astronomers to conduct a worldwide observing campaign spanning from 2023 August to 2023 December.

At the time of this study, the Unistellar fleet consisted of three telescope models: the eVscope 1, eVscope eQuinox, and eVscope 2. Each model features a 112-mm primary mirror. The eVscope 1 and eQuinox are equipped with a Sony IMX224LQR sensor, offering a 37\arcmin$\times$28\arcmin\ field of view and a pixel scale of 1.72\arcsec\ per pixel. In contrast, the eVscope 2 utilizes a Sony IMX347LQR sensor, which provides a wider field of view of 45\arcmin$\times$34\arcmin\ and a finer pixel scale of 1.33\arcsec\ per pixel. All detectors use RGB photosensors arranged in a Bayer matrix. Observations were conducted with 4.0-second exposure times, and images were taken consecutively over intervals of 20–40 minutes. AFA citizen astronomers used their personal telescopes to conduct additional observations. To ensure consistency between the AFA and Unistellar datasets, no filters were applied in any observations. As a result, the effective band passes are most comparable to the Gaia G filter or a generic “clear” filter across the different telescopes.

In total, Unistellar citizen astronomers contributed 64 successfully analyzed observations obtained from 59 different eVscopes in its global network, while AFA astronomers contributed 137 observations from 45 different telescopes.

\subsection{Reduction and Photometry} \label{sec:photometry}

AFA data were analyzed by participating astronomers with AstroLab Stellar \citep{schoolsobservatory2024}, which was utilized to conduct aperture photometry. Aperture sizes were chosen such that they encapsulate as much coma as could be detected above the background noise to measure the total flux of the comet. Video and printed tutorials were provided to the participants \citep{afastronomie_hartley2}, as well as subsequent directions to deal with background stars visible inside the coma of the comet. Brighter stars were asked to be kept outside of the aperture. The comet magnitude was then computed as the mean of the three measurements each obtained with one reference star. Error bars were reported for 83 measurements and were dominated by the standard error between reference star measurements. 

Through Unistellar's partnership with the SETI Institute, we performed photometric measurements on each eVscope observation using SETI’s automated data reduction pipeline. The images were dark-subtracted and stacked before conducting photometry. 

\begin{figure}[h]
    \centering
    \includegraphics[width=1\textwidth]{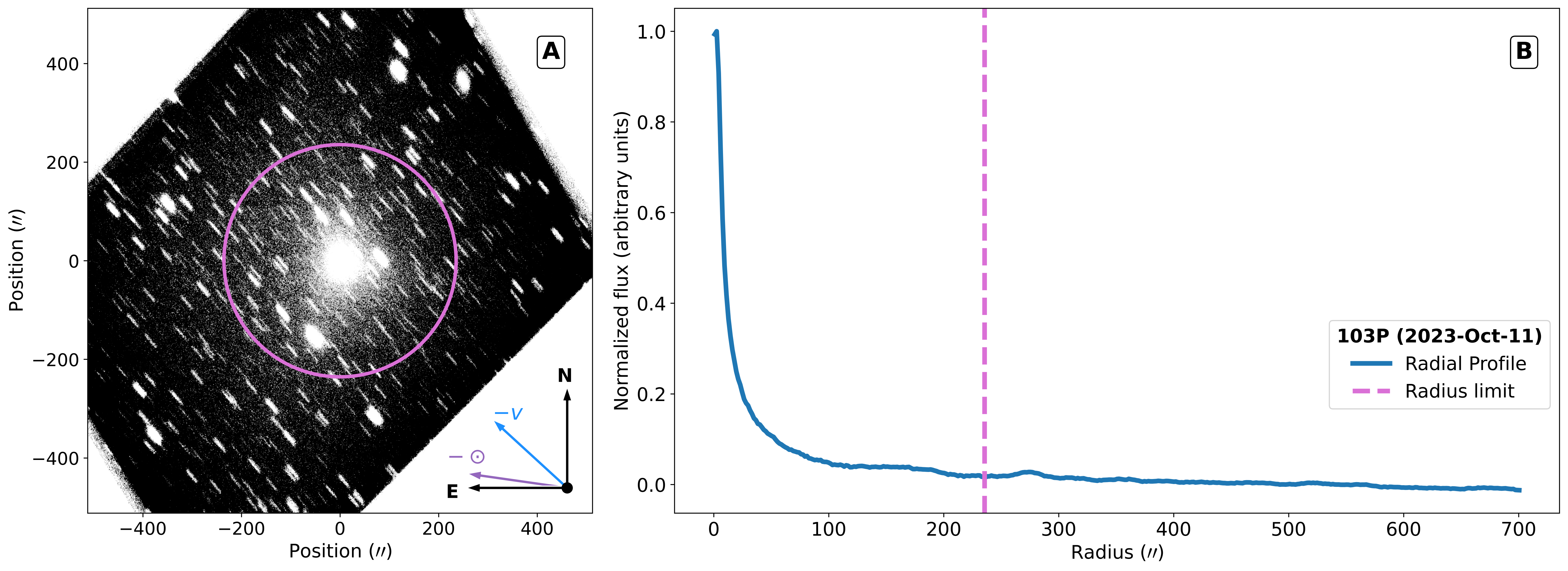}  
    \caption{Panel A shows a 20-minute stack showing the diffuse coma of Hartley 2 in a crowded star field during its 2023 apparition. Observation taken with a Unistellar eVscope. Panel B shows the surface brightness profile of the comet in blue, and the point at which the profile blends into the background of the image represented by the dashed purple line. }
    \label{fig:coma_size}
\end{figure}

Hartley 2 was in a particularly crowded field during this apparition, with a large, diffuse extended coma (see Figure~\ref{fig:coma_size}). For Unistellar data, to accurately measure the flux from Hartley 2 while minimizing contamination from background stars, we analyzed its surface brightness profile, \( S(r) \), and integrated the flux over this profile. Assuming a radially symmetric coma, the total flux, \( F_{\text{G}} \), was measured as:

\begin{equation}
    F_{\text{G}} = 2\pi \int_0^{r_{\text{max}}} S(r) \, r \, dr
\label{eq:flux}
\end{equation}

where \(r\) is the radial distance from the photocenter and \(r_{\max}\) is the maximum radius used in the integration. While Eq.~\ref{eq:flux} assumes a radially symmetric coma, we acknowledge that Hartley 2’s coma exhibits asymmetric features such as jets. However, given the relatively low resolution of our data and our use of large apertures that encompass the full coma, we expect the impact of these asymmetries on the integrated flux to be negligible. To construct the surface–brightness profile \(S(r)\), we centered on the photocenter and measured the azimuthal median flux in concentric, 1-pixel–wide annuli. We integrated only to a finite radius, $r_{\max}=\min\!\left(r_{\mathrm{spike}},\, r_{\mathrm{blend}}\right)$,
where \(r_{\mathrm{blend}}\) is the radius where the comet signal blends into the background, defined as the first annulus with \(S(r)\le B+1\sigma_B\). Here \(B\) and \(\sigma_B\) are the mean sky level and its standard deviation measured in an outer, coma-free sky annulus beyond the visible coma. Star-contamination spikes were identified directly in the radial profile using a robust local baseline \(\tilde{S}(r)\) given by a five-annulus running median. We set \(r_{\mathrm{spike}}\) to the first radius at which the excess above this baseline exceeds three times the background noise, $S(r)-\tilde{S}(r)\ \ge\ 3\,\sigma_B$, for two or more consecutive annuli.

For all measurements, the total flux was then converted into a total apparent magnitude, \( m_{\text{G}} \), using differential photometry with a set of 3–7 reference stars of similar magnitude to the expected brightness of the comet with the following

\begin{equation}
    m_{\text{G}} = -2.5 \log_{10} \left( \frac{F_{\text{G}}}{F_{\text{ref}}} \right) + m_{\text{ref}}
\end{equation}

where \( F_{\text{ref}} \) is the measured flux of the reference star and \( m_{\text{ref}} \) is the average of the known apparent magnitudes of the reference stars. Reference-star fluxes were measured by integrating their surface-brightness profiles to a radius $r_{\max}=2\times\mathrm{FWHM}$. For background subtraction, we used a local annulus measuring from $r_{\max}$ to $r_{\max}+10$ pixels. reported errors were dominated by the standard error between reference stars. A complete list of measured magnitudes and observation information is available in Table \ref{tab:photometric_data} in Appendix~\ref{sec:appendix_data}. 

\newpage

\section{Results} \label{sec:results}

\begin{figure}[h]
    \centering
    \includegraphics[width=1\linewidth]{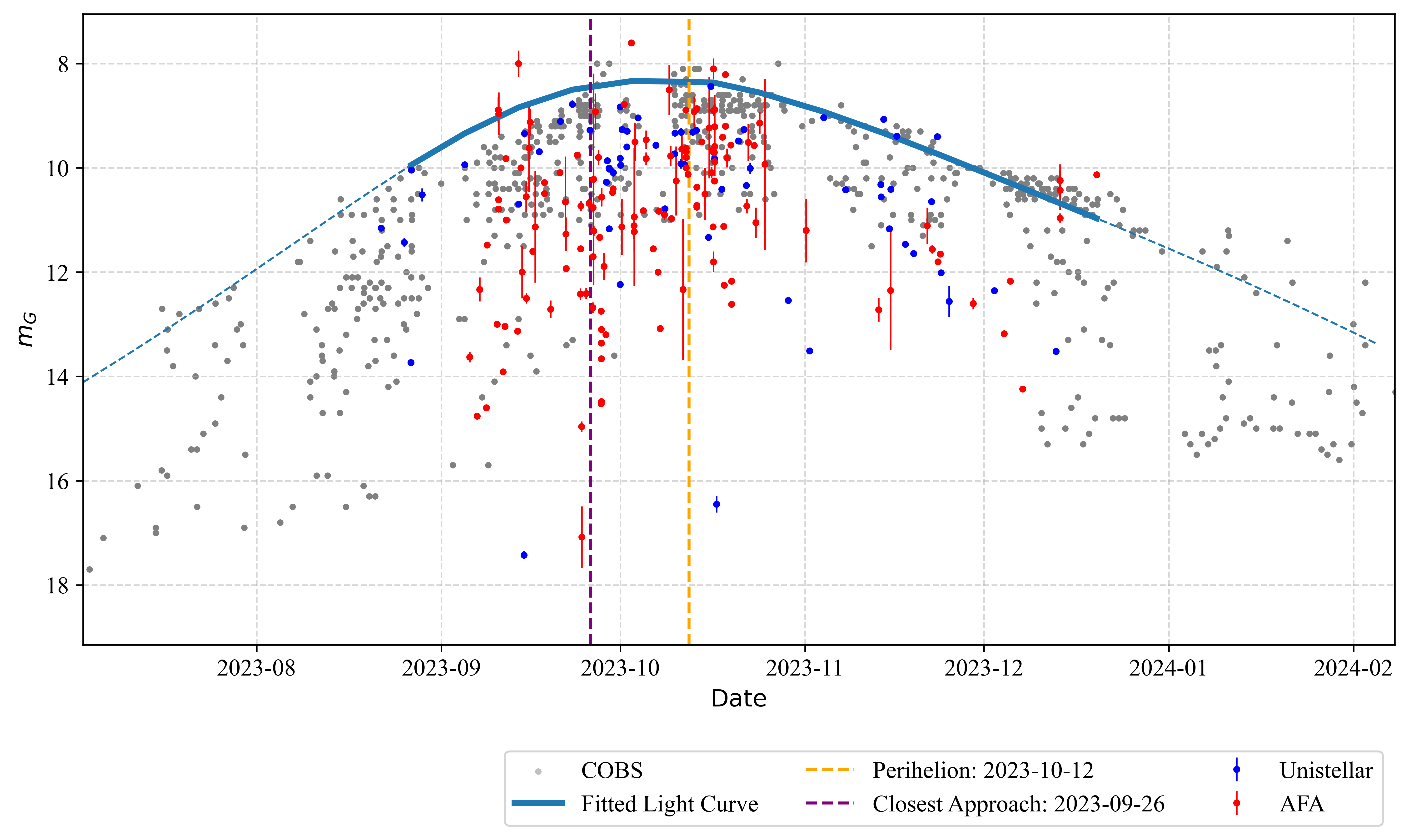}
    \caption{Lightcurve of Comet 103P/Hartley 2. The lightcurve is fitted to the upper envelope of Unistellar (blue) and AFA (red) data points. COBS submitted data is displayed in gray. The upper envelope of both sets of data points are in good agreement.}
    \label{fig:lc}
\end{figure}

The lightcurve of Hartley 2 during its 2023 apparition, constructed from our Unistellar and AFA observations, is shown in Figure~\ref{fig:lc}. There is some obvious scatter within the data points, which is consistent with similar datasets obtained from the Comet Observation Database \citep{COBS}, which also exhibit photometric dispersion. To contextualize our measurements, we overlay the COBS data, allowing direct comparison with independently collected observations.

Despite the scatter, the lightcurve reveals a clearly defined upper envelope. The brightest magnitudes likely reflect the comet's true apparent brightness under optimal observing conditions, likely when background contamination was minimal and observing conditions were favorable.

Fainter data points scattered below this upper envelope are likely impacted by several observational effects:
\begin{itemize}
    \item Background star contamination: The crowded stellar field during this apparition may have led to overlapping sources affecting flux extraction.
    \item Atmospheric conditions: Poor seeing, clouds, or high airmass could reduce the signal-to-noise ratio and artificially lower the measured brightness.
    \item Instrumental or observational effects: Differences in telescope models, tracking performance, or observer procedures may contribute to variability.
\end{itemize}

To extract a robust characterization of Hartley~2’s secular trend, we focus on the brightest magnitudes as a proxy for the comet’s intrinsic brightness. The full dataset was divided into uniform 9-day bins relative to perihelion, \((t - T_p)\), with the minimum magnitude in each bin used as our representative value. This bin size was selected to match the largest temporal gap in our Unistellar+AFA dataset. This approach robustly suppresses downward outliers caused by background interference or suboptimal observations.

To model the brightness evolution as a function of heliocentric distance \( r_h \), we fit a power-law magnitude function of the form:

\begin{equation} \label{eq:comet_magnitude}
    m_{\text{G}} = H_G + 5 \log_{10} (r_h) + k \log_{10} (\Delta),
\end{equation}

where \( m_{\text{comet}} \) is the apparent magnitude of the comet, \( H_G \) is a normalization constant, \( r_h \) is the heliocentric distance (in AU), \( \Delta \) is the geocentric distance (in AU), \( k \) is a parameter that accounts for observational geometry and phase effects. We do not apply a phase angle correction to the coma brightness, as the unfiltered observations make the relative contributions of dust and gas uncertain. As a result, the effect of dust phase scattering on the apparent brightness cannot be reliably quantified, though it is expected to be small given the modest variation in phase angle during the 2023 apparition. As observed in past apparitions, the lightcurve is asymmetrical, appearing less steep during Hartley 2's recession from perihelion than its approach to perihelion. To account for this we allow \( k \) to vary pre- and post-perihelion, which controls the fit's slope. As seen in Figure~\ref{fig:lc}, we apply this fit to only the Unistellar+AFA dataset from 2023, plotting the COBS dataset as well for comparison, with which the fit matches well. We obtain lightcurve parameters \( H_G \) = 9.85 $\pm$ 0.47, \( k_{pre} \) = 17.89 $\pm$ 9.24, and \( k_{post} \) = 14.69 $\pm$ 5.52. 

\section{Discussion} \label{sec:discussion}

\subsection{Comparison to Previous Apparitions}

\begin{figure}[h]
    \centering
    \includegraphics[width=0.8\linewidth]{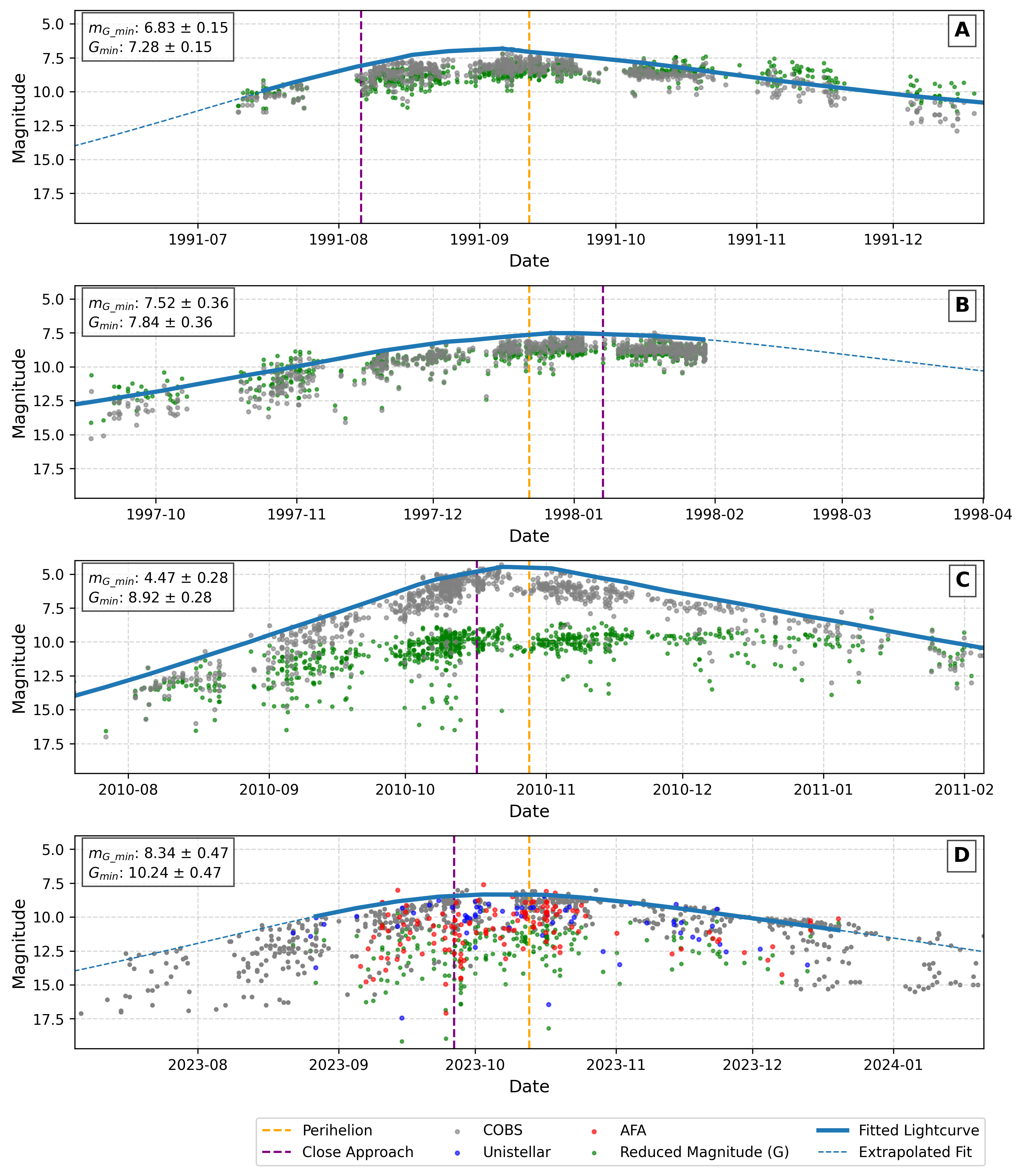}
    \caption{Lightcurves of Comet 103P/Hartley 2 from each observable apparition since 1991. Panels A–C show the fitted lightcurve (blue) over COBS data (gray). Panel D shows the lightcurve (blue) fitted only to AFA and Unistellar data (red and blue respectively), which aligns well with COBS data (gray). In all panels, reduced magnitudes are shown in green. The minimum apparent magnitude, $m_{G_{min}}$, is derived from the fit, and $G_{min}$ is the corresponding reduced magnitude. The greater scatter and uncertainties in 2023 reflect the crowded star field during that apparition.}
    \label{fig:lightcurves}
\end{figure}

To assess the ongoing secular decline in activity, we compare our fitted lightcurve to data from previous apparitions using photometry from the COBS database. Each lightcurve is modeled using Equation~\ref{eq:comet_magnitude} applying the same methodology as described in Section~\ref{sec:results}. In all cases, k is steeper pre-perihelion. Fitted parameters are listed in Table~\ref{tab:fitted_lightcurve} and fitted lightcurves are plotted in Figure~\ref{fig:lightcurves}. Lightcurve fitting only utilizes the date range (and corresponding heliocentric/geocentric distances) covered by the available dataset, shown as solid blue lines in Figure~\ref{fig:lightcurves}. Dashed lines are extrapolations of the fits. For the 1991, 1997, and 2010 apparitions, we fit only COBS data (Panels A–C), while the 2023 fit uses only the Unistellar+AFA dataset (Panel D), though COBS data are included for comparison. The date ranges are provided in Table~\ref{tab:fitted_lightcurve}. Uncertainties in the fitted values of $H_G$, $k_{\text{pre}}$, and $k_{\text{post}}$ were estimated as the square root of the diagonal elements of the covariance matrix returned by the \texttt{curve\_fit} routine from the \texttt{scipy.optimize} Python package, corresponding to formal $1\sigma$ uncertainties.

\begin{table}[htbp]
    \caption{Fitted lightcurve parameters.}
    \label{tab:fitted_lightcurve}
    \begin{tabular}{lcccc}
        \hline \hline
        Apparition & $H_G$ & $k$ (Pre-perihelion) & $k$ (Post-perihelion) & Date Range (UTC) \\
        \hline
        1991 & 7.61 $\pm$ 0.15 & 25.60 $\pm$ 3.98 & 12.73 $\pm$ 1.25 & 1991 Jul 9–1992 Jan 12\\
        1997 & 7.83 $\pm$ 0.36 & 21.83 $\pm$ 1.56 & 6.83 $\pm$ 13.26 & 1997 May 3–1998 Jan 29\\
        2010 & 8.40 $\pm$ 0.28 & 26.72 $\pm$ 1.81 & 11.52 $\pm$ 1.21 & 2010 May 13–2011 Jun 4\\
        2023 (This work) & 9.85 $\pm$ 0.47 & 17.89 $\pm$ 9.24 & 14.69 $\pm$ 5.52 &  2023 Aug 21–2023 Dec 19\\
        \hline
    \end{tabular}
\end{table}

To compare activity between each apparition, we normalize the peak magnitude from each fit to represent the apparent brightness if the comet was placed at a heliocentric distance, \( r_h \), and geocentric distance, \( \Delta \), of 1 AU. This reduced magnitude, \( G \), is calculated with:

\begin{equation}
    G = m_{G} - 5 \log_{10} (r_h \Delta)
\end{equation}

 This adjustment allows for more direct comparisons across different apparitions. The minimum magnitude value derived from each apparition shows a clear decreasing trend, indicating that the comet is becoming progressively fainter over successive perihelion passages. Table~\ref{tab:perihelion_data} summarizes the minimum measured magnitude for each past apparition, alongside the corresponding heliocentric and geocentric distances at perihelion. As previously noted by \citet{knight2013} and \citet{schleicher2007}, the observed decline in brightness exceeds what would be expected from changes in heliocentric and geocentric distances alone. This conclusion is further supported by the consistent increase in both the normalization constant, $H_G$, and the minimum reduced magnitude, $G_{min}$.

\begin{table*}[h]
\centering
\caption{Comet 103P/Hartley 2 at perihelion.}
\label{tab:perihelion_data}

\begin{threeparttable}
\begin{tabularx}{0.65\textwidth}{lccccc}
\hline
Perihelion Date     & $r_h$ (AU)\tnote{a} & $\Delta$ (AU)\tnote{b} &
Peak Date           & $m_{G\_\mathrm{min}}\tnote{c}$ & $G_\mathrm{min}\tnote{d}$ \\
\hline
1991 Sep 11 & 0.95 & 0.719 & 1991 Sep 9 & $6.83 \pm 0.15$ & $7.28 \pm 0.15$\\
1997 Dec 22 & 1.03 & 0.700 & 1997 Dec 26 & $7.52 \pm 0.36$ & $7.84 \pm 0.36$\\
2004 May 17 & 1.04 & 0.700 & — & — & —\\
2010 Oct 28 & 1.06 & 0.695 & 2010 Oct 22 & $4.47 \pm 0.28$ & $8.92 \pm 0.28$\\
2017 Apr 20 & 1.07 & 0.693 & — & — & —\\
2023 Oct 12 & 1.06 & 0.693 & 2023 Oct 2 & $8.34 \pm 0.47$ & $10.24 \pm 0.47$\\
\hline
\end{tabularx}

\begin{tablenotes}
\item[a] Heliocentric distance
\item[b] Geocentric distance
\item[c] Apparent magnitude
\item[d] Reduced magnitude
\end{tablenotes}
\end{threeparttable}
\end{table*}

\subsection{Long-Term Fate of Comet 103P/Hartley 2}\label{Long-Term Fate of 103P/Hartley 2}

\begin{figure}[h]
    \centering
    \includegraphics[width=1\linewidth]{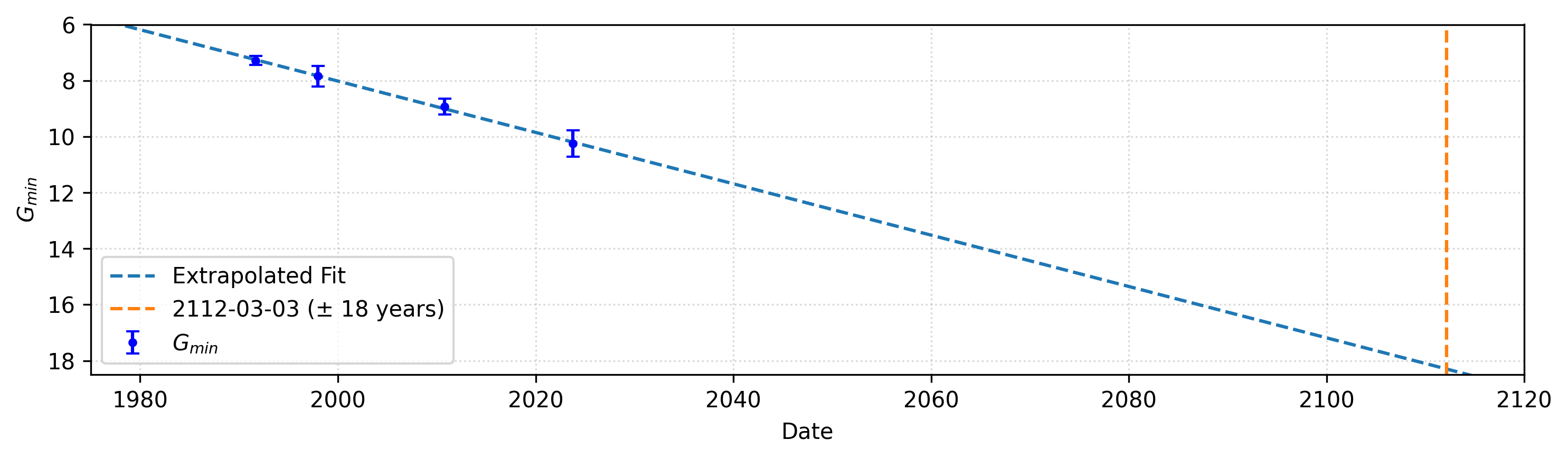}
    \caption{The observed decline in minimum reduced magnitude of Comet 103P/Hartley 2 across multiple apparitions. Previous apparitions include only COBS data, the 2023 apparition includes data from this work as well. The decreasing trend in brightness is indicative of progressive volatile depletion rather than changes in orbital geometry. If this trend continues, at the earliest, Hartley 2 will reach nuclear magnitude in year$\sim$2112, after $\sim$13 more active apparitions.}
    \label{fig:decay}
\end{figure}

When the minimum reduced magnitude, $G_{min}$, of each apparition is plotted over time, the trend of activity decline becomes clear (Fig. \ref{fig:decay}). The steady fading has indeed continued through the comet's latest apparition in 2023. Our results are consistent with recent narrow-band photometry presented by \citet{schleicher2024} which also found that the long-term secular decrease in production rates has persisted through the 2023 apparition. Fitting a linear trend shows that $G_{min}$ decreases by 0.59 $\pm$ 0.11 every apparition (6.48 years). If this activity decline continues, eventually only the nucleus will contribute to the comet’s brightness.

To estimate the absolute magnitude of Hartley 2's nucleus in the \textit{Gaia} $G$ band, we convert from its reported absolute $V$-band magnitude using the color transformation presented in \citet{jordi2010}. Given a $V$-band absolute magnitude of $18.4 \pm 0.1$ and color indices $B - V = 0.75 \pm 0.05$ and $V - R = 0.43 \pm 0.04$ \citep{li2013}, we apply the following empirical relation:

\begin{equation}
G = V - 0.0176 - 0.00686(B - V) - 0.1732(V - R)
\end{equation}

Substituting the known values, we obtain an absolute $G$-band magnitude of $G = 18.30 \pm 0.10$. If Hartley 2’s decaying activity trend continues as shown in Figure~\ref{fig:decay}, the comet will reach nuclear magnitude $V = 18.4 \pm 0.1$ ($G = 18.3 \pm 0.1$) in the year 2112 $\pm$ 18. Assuming that, by this time, all observed flux originates solely from the nucleus due to the cessation of activity, this timeline implies approximately 13 more active orbits remain (considering an orbital period of 6.47 years). However, the comet's activity is most likely to decline asymptotically, remaining weakly active for much longer than this estimation. Thus, 13 orbits should be considered as a lower limit.

Interpreting the secular fading in terms of flux, a decline of $\Delta m = 0.59\pm0.11$~mag per orbit corresponds to a per‐orbit flux ratio $f = 10^{-0.4\Delta m} = 0.58\pm0.06$, i.e., an average decrease of $42\%\pm6\%$ in activity each return. Compounded over the $\sim$5 orbits between 1991 and 2023, this implies a median flux ratio of $(0.58)^5 \simeq 0.065$, meaning Hartley 2’s peak activity is now lower by a factor of $\sim$15 compared to 1991 (with a $1\sigma$ range of $\sim$9–26 when propagating the uncertainty in $\Delta m$). If the present trend persists, the comet would be less than $\sim\!0.01\%$ of the 1991 activity level after 13 further apparitions from 2023, fading asymptotically rather than ceasing entirely, with weak activity likely persisting beyond that time. \citet{knight2013} report a $\sim$40\% drop in water production per apparition (between 1991–1997 and 1997–2010). Our broadband (gas+dust) decline of $42\%\pm6\%$ per orbit is broadly consistent with those gas–only results, particularly between 1991 and 1997, though we find a steep decline between 1997 and 2010. We note that our mass‐loss estimate is an upper limit on dust-driven loss, because our unfiltered photometry likely includes a non-negligible gas contribution. Based on typical comet spectra and Hartley 2’s known low dust-to-gas ratio \citep{weaver1994,ahearn1995}, gas may account for $\sim$15–25\% of the detected flux; thus reduced magnitudes likely overestimate the dust contribution, and inferred dust mass‐loss rates should be treated as upper limits. 

With flux proportional to active fraction, this decline suggests that Hartley 2’s effective active fraction has dropped by roughly an order of magnitude since its 1991 apparition. Early estimates based on pre-EPOXI size determinations found that the active fraction of the comet's surface was $f_{\rm active} > 1$ at perihelion during the 1997 and 2010 apparitions, respectively \citep{Groussin2004, Lisse2009}. The “hyperactive” characterization became widely used after EPOXI revealed that centimeter–scale icy chunks released from CO\(_2\)–driven jets boost the coma’s water production, explaining the implausibly large surface active fractions \citep{AHearn2011,Kelley2013}. Because, to first order, the broadband coma flux serves as a proxy for production rate and $Q \propto f_{\rm active}$ for fixed nucleus area and sublimation efficiency, we take $f_{\rm active}\propto F_{\rm G}$ between apparitions, and thus we can estimate active fractions directly from the measured flux ratios. Using $f_{\rm active}\!\sim\!1.17$ at perihelion during the 2010 apparition \citep{Lisse2009}, we can estimate the active surface fraction at subsequent apparitions by applying our per-orbit activity decline of $42\%\pm6\%$ (i.e., multiplying by $f=0.58\pm0.06$ per return) and assuming a constant nucleus size across apparitions. We find $f_{\rm active}(2017)\sim 0.7$ and $f_{\rm active}(2023) \sim 0.4$. Therefore, Hartley 2 likely no longer should carry its “hyperactive’’ designation as of at least the 2017 apparition.

The long-term evolution of Hartley 2 hinges on which mechanism primarily drives its activity decline. Several processes can contribute to secular fading in Jupiter-family comets:

\begin{enumerate}
    \item Volatile depletion: Repeated perihelion passages progressively deplete near-surface volatiles, reducing the comet's ability to sustain activity \citep{levison1997, fernandez2002, gillan2024}. This process may dominate for smaller nuclei with short orbital periods, such as Hartley 2.

    \item Uneven volatile distribution: If activity arises from a localized volatile-rich patch near a rotation pole, slow precession can gradually tip that area out of sunlight, reducing sublimation. This mechanism explains the fading of 9P/Tempel 1 \citep{schleicher2007}, but current data imply it is secondary for Hartley 2 \citep{knight2013}. Complex, non-principal-axis rotation can similarly result in varying levels of surface area exposure to the Sun, and can therefore still modulate the comet’s output \citep{knight2013, knight2015}.

    \item Mantle formation: Sublimation leaves behind refractory material that can accumulate into a thermally insulating mantle, eventually suppressing activity \citep{rickman1990}. The Rosetta spacecraft showed that on comet 67P/Churyumov–Gerasimenko, seasonal dust back-fall can accumulate to decimeter–meter thickness within a single orbit, especially in the northern smooth terrains, yet such areas can remain active when illuminated, with activity reconfigured toward scarps and edges\citep{Hu2017,keller2017}. No observed decline in activity has been reported in 67P from apparition to apparition. Also, given Hartley 2's smaller size relative to 67P (mean radius 0.58 km vs. ~1.65 km), the backfall fraction may be smaller. Thus mantle growth on longer timescales could plausibly contribute to Hartley 2’s fading, though it likely does not fully explain the observed decline.

    \item Fragmentation and disintegration: Rather than crust development leading to dormancy, fragmentation is more likely to lead to disintegration before then \citep{chen1994, disisto2009, jewitt2022}. Although Hartley 2 has not undergone wholesale fragmentation, its well-documented hyperactivity is thought to arise from large, low-density icy chunks that detach from the nucleus and sublimate in the coma \citep{AHearn2011, Kelley2013}. These “mini-fragments” blur the line between surface-driven activity and fragmentation. However, fragmentation, especially catastrophic fragmentation will lead to a sudden brightening followed by a rapid decrease in active, and this has not been observed in Hartley 2, so fragmentation and disintegration cannot be attributed to the observed activity decline. 
\end{enumerate}

Typical active lifetimes for JFCs range from $\sim10^{3}$–$10^{4}$ yr after entry into the inner Solar System \citep{levison1997, fernandez2009}.  Mass-loss estimates of $10^{7}$–$10^{8}$ kg per return \citep{AHearn2011, Kelley2013, Bauer2011} imply that Hartley 2 could survive several hundred more returns, yet its current fading trend projects as low as $\sim$13 additional apparitions before activity becomes marginal. Whether the comet ultimately fragments or mantles into dormancy will depend on the balance between ongoing volatile loss and crust formation. If a dust mantle forms more rapidly than material is lost, the comet may transition into a dormant or extinct state. Many near-Earth objects (NEOs) are believed to be such evolved comets, exhibiting little or no activity despite their cometary origin \citep{weissman2002, jewitt2004, whitman2006}. Thermal models and observations suggest thick refractory mantles can suppress outgassing, allowing JFCs to persist as inert, asteroid-like bodies \citep{rahe1994, jewitt2022}. Alternatively, if mass loss remains substantial and volatiles continue to be exposed, structural weakening could lead to fragmentation, as seen in comets like 73P/Schwassmann-Wachmann 3, which broke apart in 1995 and continued shedding material in later apparitions \citep{scotti1996, reach2009, weaver2006}. Considering Hartley 2’s small nucleus, high activity, and lack of major outbursts or fragment trails, we regard progressive volatile depletion, possibly augmented by localized mantle growth and the waning supply of hyperactive chunks, as the most plausible primary driver of its secular fading.

Continued monitoring over the next few returns will be critical for constraining Hartley 2’s evolutionary trajectory. If current trends continue, Hartley 2 may soon enter its final active stages.

\section{Conclusions} \label{sec:conclusions}

Our analysis of photometric observations of Comet 103P/Hartley 2 during its 2023 apparition obtained through a broad network of citizen astronomers confirms a continued secular decline in activity. This fading is most likely driven by intrinsic physical changes in the nucleus, rather than changes in orbital geometry. The long-term trend is consistent with evolutionary processes observed in other Jupiter-family comets (JFCs), where repeated perihelion passages gradually deplete near-surface volatiles and modify surface morphology. Our principal findings are:

\begin{itemize}
    \item The lightcurve fit for the 2023 apparition yields a minimum apparent magnitude $m_{G,\mathrm{min}} = 8.34 \pm 0.47$, consistent with independently reported values from the Comet Observation Database (COBS).
    
    \item The minimum reduced magnitude, $m_{G,\mathrm{min}}$ has increased across successive apparitions, indicating a progressive decline in activity.
    
    \item The long-term brightness decline, observed since at least 1991, continues at a rate of $0.59 \pm 0.11$~mag per orbit (6.48~yr), equivalent to a per-orbit flux loss factor of $f = 10^{-0.4\Delta m} = 0.58 \pm 0.06$.
    
    \item If this fading continues, Hartley 2 is expected to remain active for at least 13 more apparitions. However, the decline is likely asymptotic, meaning weak activity could persist well beyond this lower-limit estimate.
    
    \item As brightness declines, the fraction of the nucleus surface contributing to activity ($f_{\rm active}$) is also expected to fall. Assuming $f_{\rm active} \propto$ flux, we estimate an average decrease of $42\% \pm 6\%$ in active area per orbit. Given that Hartley 2 was estimated to be hyperactive ($f_{\rm active} \sim 1.17$) at perihelion in 2010 \citep{Lisse2009}, our results suggest it likely was no longer in a hyperactive state by the 2017 apparition.
\end{itemize}

Future observations will be crucial for determining whether Hartley 2 ultimately becomes dormant or undergoes disintegration. Continued photometric and spectroscopic monitoring, particularly through the growing capabilities of coordinated citizen science networks, will further refine our understanding of mass loss, volatile depletion, and the end stages of cometary evolution. This work highlights the essential role of professional–amateur (ProAm) collaborations in tracking the evolutionary paths of comets.
\\
\\
We thank the anonymous referees for their thoughtful reviews and constructive suggestions, which improved this manuscript. This work was supported by the Moore Foundation and the Association Fran\c{c}aise d'Astronomie.

\appendix
\section{Photometric Data Table} \label{sec:appendix_data}

A complete list of our photometry of Hartley 2 during its 2023 apparition. In some cases photometric radii or errors were not reported, or were measured to be insignificantly small, and are left out of the table. 

\begin{longtable}{lccccc}

\caption{Photometric Measurements of 103P/Hartley 2 \label{tab:photometric_data}} \\
\hline
Observation Date & Apparent Magnitude ($m_G$) & Photometric Radius ($\prime$) & $r_h$ (AU) & $\Delta$ (AU) & Affiliation \\
\hline
\endfirsthead

\multicolumn{6}{c}{{\bfseries \tablename\ \thetable{} -- continued from previous page}} \\
\hline
Observation Date & Apparent Magnitude ($m_G$) & Photometric Radius ($\prime$) & $r_h$ (AU) & $\Delta$ (AU) & Affiliation \\
\hline
\endhead

\hline \multicolumn{6}{r}{{Continued on next page}} \\
\endfoot

\hline
\endlastfoot

2023-Aug-21 21:22:49.45 & $11.15 \pm 0.01$ & 14.02 & 1.27 & 0.52 & Unistellar \\
2023-Aug-25 18:02:06.35 & $11.43 \pm 0.08$ & 12.60 & 1.25 & 0.49 & Unistellar \\
2023-Aug-26 21:17:50.41 & $13.73 \pm 0.03$ & 12.12 & 1.24 & 0.49 & Unistellar \\
2023-Aug-26 22:57:22.99 & $10.04 \pm 0.01$ & 1.06 & 1.24 & 0.48 & Unistellar \\
2023-Aug-28 18:12:55.14 & $10.52 \pm 0.12$ & 11.41 & 1.23 & 0.47 & Unistellar \\
2023-Sep-04 21:58:49.35 & $9.94 \pm 0.03$ & 2.79 & 1.18 & 0.43 & Unistellar \\
2023-Sep-05 17:45:13.00 & $13.63 \pm 0.10$ & 0.63 & 1.18 & 0.43 & AFA \\
2023-Sep-06 23:34:35.99 & $14.76$ & -- & 1.17 & 0.42 & AFA \\
2023-Sep-07 10:10:01.00 & $12.33 \pm 0.23$ & 1.40 & 1.17 & 0.42 & AFA \\
2023-Sep-08 13:08:21.99 & $14.60$ & -- & 1.16 & 0.42 & AFA \\
2023-Sep-08 15:25:39.99 & $11.48$ & -- & 1.16 & 0.42 & AFA \\
2023-Sep-10 07:44:46.99 & $13.00$ & -- & 1.15 & 0.41 & AFA \\
2023-Sep-10 13:03:09.00 & $8.89 \pm 0.25$ & 7.16 & 1.15 & 0.41 & AFA \\
2023-Sep-10 13:46:58.99 & $10.79 \pm 0.01$ & 4.16 & 1.15 & 0.41 & AFA \\
2023-Sep-10 13:51:28.00 & $10.61 \pm 0.02$ & 4.17 & 1.15 & 0.41 & AFA \\
2023-Sep-10 14:26:51.99 & $8.96 \pm 0.41$ & 6.36 & 1.15 & 0.41 & AFA \\
2023-Sep-11 07:47:00.00 & $13.91$ & -- & 1.15 & 0.41 & AFA \\
2023-Sep-11 15:23:43.99 & $13.04$ & -- & 1.15 & 0.41 & AFA \\
2023-Sep-11 19:21:04.00 & $9.82$ & -- & 1.14 & 0.41 & AFA \\
2023-Sep-11 20:47:53.00 & $11.00$ & -- & 1.14 & 0.41 & AFA \\
2023-Sep-13 18:21:04.99 & $13.13 \pm 0.02$ & -- & 1.14 & 0.40 & AFA \\
2023-Sep-13 22:13:57.00 & $8.00 \pm 0.25$ & -- & 1.13 & 0.40 & AFA \\
2023-Sep-13 23:10:51.26 & $10.69 \pm 0.02$ & 17.57 & 1.13 & 0.40 & Unistellar \\
2023-Sep-14 07:54:08.99 & $10.00$ & -- & 1.13 & 0.40 & AFA \\
2023-Sep-14 12:15:47.00 & $12.00 \pm 0.50$ & -- & 1.13 & 0.40 & AFA \\
2023-Sep-14 19:51:57.88 & $17.43 \pm 0.08$ & 15.13 & 1.13 & 0.40 & Unistellar \\
2023-Sep-14 22:02:08.64 & $9.34 \pm 0.07$ & 0.09 & 1.13 & 0.40 & Unistellar \\
2023-Sep-15 04:56:01.99 & $10.55 \pm 0.29$ & 4.81 & 1.13 & 0.40 & AFA \\
2023-Sep-15 05:54:26.00 & $12.50 \pm 0.10$ & -- & 1.13 & 0.40 & AFA \\
2023-Sep-15 16:26:09.99 & $9.62 \pm 0.84$ & 2.68 & 1.13 & 0.39 & AFA \\
2023-Sep-15 20:54:04.99 & $9.12 \pm 0.26$ & 5.53 & 1.13 & 0.39 & AFA \\
2023-Sep-16 07:53:36.00 & $11.60$ & -- & 1.12 & 0.39 & AFA \\
2023-Sep-16 17:06:39.00 & $11.13 \pm 1.07$ & 4.04 & 1.12 & 0.39 & AFA \\
2023-Sep-17 09:37:16.73 & $9.69 \pm 0.03$ & 22.09 & 1.12 & 0.39 & Unistellar \\
2023-Sep-18 06:47:02.99 & $10.49 \pm 0.01$ & 4.39 & 1.12 & 0.39 & AFA \\
2023-Sep-18 06:49:47.00 & $10.28 \pm 0.03$ & 4.62 & 1.12 & 0.39 & AFA \\
2023-Sep-19 07:15:38.00 & $12.71 \pm 0.17$ & 1.05 & 1.11 & 0.39 & AFA \\
2023-Sep-20 20:36:38.99 & $10.09 \pm 0.01$ & 3.76 & 1.11 & 0.39 & AFA \\
2023-Sep-20 22:34:59.93 & $9.11 \pm 0.05$ & 1.60 & 1.11 & 0.39 & Unistellar \\
2023-Sep-21 18:43:46.99 & $10.65 \pm 0.87$ & 3.02 & 1.10 & 0.38 & AFA \\
2023-Sep-21 21:05:24.99 & $11.27 \pm 0.32$ & 4.04 & 1.10 & 0.38 & AFA \\
2023-Sep-21 21:37:43.00 & $11.93 \pm 0.03$ & -- & 1.10 & 0.38 & AFA \\
2023-Sep-22 22:40:48.59 & $8.78 \pm 0.08$ & 13.05 & 1.10 & 0.38 & Unistellar \\
2023-Sep-23 18:36:12.00 & $9.75$ & -- & 1.10 & 0.38 & AFA \\
2023-Sep-24 07:35:09.00 & $12.42 \pm 0.11$ & 1.05 & 1.09 & 0.38 & AFA \\
2023-Sep-24 08:24:53.99 & $11.55$ & -- & 1.09 & 0.38 & AFA \\
2023-Sep-24 09:07:02.99 & $10.73 \pm 0.09$ & 2.33 & 1.09 & 0.38 & AFA \\
2023-Sep-24 12:16:39.99 & $14.96 \pm 0.10$ & -- & 1.09 & 0.38 & AFA \\
2023-Sep-24 12:51:49.99 & $17.08 \pm 0.59$ & -- & 1.09 & 0.38 & AFA \\
2023-Sep-25 06:45:58.00 & $12.41 \pm 0.11$ & 1.05 & 1.09 & 0.38 & AFA \\
2023-Sep-25 18:14:29.99 & $10.68 \pm 0.09$ & 4.61 & 1.09 & 0.38 & AFA \\
2023-Sep-25 22:09:06.30 & $9.27 \pm 0.01$ & 12.26 & 1.09 & 0.38 & Unistellar \\
2023-Sep-26 09:24:15.00 & $12.68 \pm 0.10$ & 0.74 & 1.09 & 0.38 & AFA \\
2023-Sep-26 10:29:42.00 & $10.76 \pm 0.02$ & 2.23 & 1.09 & 0.38 & AFA \\
2023-Sep-26 10:34:37.99 & $11.70$ & -- & 1.09 & 0.38 & AFA \\
2023-Sep-26 12:33:47.99 & $10.22 \pm 2.03$ & 2.31 & 1.09 & 0.38 & AFA \\
2023-Sep-26 12:33:47.99 & $10.22 \pm 2.03$ & 2.31 & 1.09 & 0.38 & AFA \\
2023-Sep-26 12:47:05.00 & $11.21 \pm 0.10$ & -- & 1.09 & 0.38 & AFA \\
2023-Sep-26 19:46:51.00 & $8.92 \pm 0.35$ & 5.63 & 1.09 & 0.38 & AFA \\
2023-Sep-27 07:52:46.00 & $9.80 \pm 0.15$ & 5.27 & 1.08 & 0.38 & AFA \\
2023-Sep-27 13:17:28.99 & $11.33$ & -- & 1.08 & 0.38 & AFA \\
2023-Sep-27 18:56:07.00 & $14.52$ & -- & 1.08 & 0.38 & AFA \\
2023-Sep-27 19:08:23.99 & $12.75$ & -- & 1.08 & 0.38 & AFA \\
2023-Sep-27 19:38:20.99 & $13.10$ & -- & 1.08 & 0.38 & AFA \\
2023-Sep-27 19:49:50.99 & $13.36$ & -- & 1.08 & 0.38 & AFA \\
2023-Sep-27 19:57:26.00 & $13.66$ & -- & 1.08 & 0.38 & AFA \\
2023-Sep-27 20:09:04.99 & $14.48$ & -- & 1.08 & 0.38 & AFA \\
2023-Sep-27 21:10:08.00 & $10.56 \pm 0.18$ & 5.94 & 1.08 & 0.38 & AFA \\
2023-Sep-28 06:24:56.00 & $11.89 \pm 0.26$ & 1.41 & 1.08 & 0.38 & AFA \\
2023-Sep-28 14:25:23.00 & $13.20$ & 20.00 & 1.08 & 0.38 & AFA \\
2023-Sep-28 16:19:43.22 & $10.27 \pm 0.06$ & 10.39 & 1.08 & 0.38 & Unistellar \\
2023-Sep-28 19:42:04.16 & $9.86 \pm 0.02$ & 16.08 & 1.08 & 0.38 & Unistellar \\
2023-Sep-29 03:00:07.47 & $11.17 \pm 0.03$ & 17.36 & 1.08 & 0.38 & Unistellar \\
2023-Sep-29 03:43:41.79 & $10.02 \pm 0.02$ & 0.93 & 1.08 & 0.38 & Unistellar \\
2023-Sep-29 03:46:31.40 & $10.00 \pm 0.01$ & 3.17 & 1.08 & 0.38 & Unistellar \\
2023-Sep-29 16:42:43.99 & $10.47 \pm 0.06$ & 4.13 & 1.08 & 0.38 & AFA \\
2023-Sep-29 18:25:48.00 & $10.40 \pm 0.05$ & 3.93 & 1.08 & 0.38 & AFA \\
2023-Sep-29 19:49:14.42 & $10.09 \pm 0.08$ & 4.57 & 1.08 & 0.38 & Unistellar \\
2023-Sep-30 23:24:50.85 & $8.83 \pm 0.03$ & 12.29 & 1.08 & 0.39 & Unistellar \\
2023-Sep-30 23:44:38.60 & $9.82 \pm 0.05$ & 10.02 & 1.08 & 0.39 & Unistellar \\
2023-Sep-30 23:51:59.19 & $12.24 \pm 0.04$ & 18.64 & 1.08 & 0.39 & Unistellar \\
2023-Oct-01 01:00:03.57 & $9.95 \pm 0.02$ & 4.37 & 1.08 & 0.39 & Unistellar \\
2023-Oct-01 06:13:46.00 & $11.13 \pm 0.54$ & -- & 1.08 & 0.39 & AFA \\
2023-Oct-01 07:15:50.86 & $9.26 \pm 0.07$ & 3.43 & 1.08 & 0.39 & Unistellar \\
2023-Oct-01 15:39:04.99 & $8.78$ & 8.30 & 1.07 & 0.39 & AFA \\
2023-Oct-02 02:27:58.89 & $9.60 \pm 0.01$ & 18.01 & 1.07 & 0.39 & Unistellar \\
2023-Oct-02 03:28:31.59 & $9.30 \pm 0.03$ & 23.60 & 1.07 & 0.39 & Unistellar \\
2023-Oct-02 20:26:49.00 & $7.60$ & 1.26 & 1.07 & 0.39 & AFA \\
2023-Oct-03 07:15:11.00 & $11.11 \pm 0.25$ & 3.74 & 1.07 & 0.39 & AFA \\
2023-Oct-03 07:44:04.99 & $11.22 \pm 0.37$ & 3.60 & 1.07 & 0.39 & AFA \\
2023-Oct-03 07:50:14.00 & $10.94 \pm 1.32$ & 2.63 & 1.07 & 0.39 & AFA \\
2023-Oct-03 11:17:31.00 & $9.50 \pm 0.36$ & 9.53 & 1.07 & 0.39 & AFA \\
2023-Oct-03 23:29:49.19 & $9.04 \pm 0.02$ & 20.00 & 1.07 & 0.39 & Unistellar \\
2023-Oct-04 20:11:33.99 & $10.82$ & -- & 1.07 & 0.39 & AFA \\
2023-Oct-05 08:18:06.00 & $9.82 \pm 0.12$ & 2.44 & 1.07 & 0.39 & AFA \\
2023-Oct-05 08:23:45.99 & $9.46 \pm 0.18$ & 2.42 & 1.07 & 0.39 & AFA \\
2023-Oct-06 12:21:53.99 & $11.55 \pm 0.04$ & 3.22 & 1.07 & 0.39 & AFA \\
2023-Oct-06 23:35:44.17 & $9.56 \pm 0.06$ & 15.89 & 1.07 & 0.39 & Unistellar \\
2023-Oct-07 08:18:28.99 & $12.00$ & -- & 1.07 & 0.39 & AFA \\
2023-Oct-07 10:51:57.00 & $10.83$ & 0.61 & 1.07 & 0.39 & AFA \\
2023-Oct-07 16:24:35.99 & $13.08 \pm 0.01$ & 1.00 & 1.07 & 0.40 & AFA \\
2023-Oct-08 09:48:06.00 & $10.90 \pm 0.10$ & 0.50 & 1.07 & 0.40 & AFA \\
2023-Oct-08 11:50:12.53 & $10.78 \pm 0.01$ & 11.56 & 1.07 & 0.40 & Unistellar \\
2023-Oct-09 04:56:17.00 & $8.50 \pm 0.48$ & -- & 1.07 & 0.40 & AFA \\
2023-Oct-09 10:34:27.00 & $9.77 \pm 0.19$ & 9.02 & 1.06 & 0.40 & AFA \\
2023-Oct-09 14:13:51.00 & $10.97$ & -- & 1.06 & 0.40 & AFA \\
2023-Oct-10 03:14:37.21 & $9.73 \pm 0.02$ & 20.59 & 1.06 & 0.40 & Unistellar \\
2023-Oct-10 04:19:41.33 & $9.33 \pm 0.01$ & 5.05 & 1.06 & 0.40 & Unistellar \\
2023-Oct-10 08:38:36.99 & $10.25 \pm 0.66$ & 18.33 & 1.06 & 0.40 & AFA \\
2023-Oct-11 03:54:24.83 & $9.92 \pm 0.09$ & 7.10 & 1.06 & 0.40 & Unistellar \\
2023-Oct-11 04:57:40.22 & $9.31 \pm 0.08$ & 6.17 & 1.06 & 0.40 & Unistellar \\
2023-Oct-11 07:50:02.99 & $9.64 \pm 0.33$ & 11.07 & 1.06 & 0.40 & AFA \\
2023-Oct-11 12:36:49.99 & $12.33 \pm 1.35$ & 1.20 & 1.06 & 0.40 & AFA \\
2023-Oct-11 19:34:53.08 & $9.93 \pm 0.01$ & 0.09 & 1.06 & 0.40 & Unistellar \\
2023-Oct-12 00:22:40.99 & $8.89 \pm 0.06$ & 2.64 & 1.06 & 0.40 & AFA \\
2023-Oct-12 00:39:26.99 & $9.62$ & 4.12 & 1.06 & 0.40 & AFA \\
2023-Oct-12 00:43:26.00 & $9.80$ & 3.65 & 1.06 & 0.40 & AFA \\
2023-Oct-12 00:46:35.00 & $9.69$ & 4.62 & 1.06 & 0.40 & AFA \\
2023-Oct-12 00:48:58.99 & $10.00$ & 3.51 & 1.06 & 0.40 & AFA \\
2023-Oct-12 08:41:42.00 & $10.12$ & 2.64 & 1.06 & 0.41 & AFA \\
2023-Oct-13 03:49:21.81 & $9.31 \pm 0.01$ & 2.52 & 1.06 & 0.41 & Unistellar \\
2023-Oct-13 09:12:36.00 & $8.92 \pm 0.29$ & 11.00 & 1.06 & 0.41 & AFA \\
2023-Oct-13 18:19:43.36 & $9.28 \pm 0.03$ & 9.17 & 1.06 & 0.41 & Unistellar \\
2023-Oct-13 19:45:28.99 & $10.72$ & 2.64 & 1.06 & 0.41 & AFA \\
2023-Oct-13 19:47:17.00 & $10.75$ & 2.64 & 1.06 & 0.41 & AFA \\
2023-Oct-13 19:48:53.00 & $10.37$ & 2.64 & 1.06 & 0.41 & AFA \\
2023-Oct-13 19:52:44.99 & $8.86$ & 4.12 & 1.06 & 0.41 & AFA \\
2023-Oct-14 16:02:33.00 & $9.50$ & 0.48 & 1.06 & 0.41 & AFA \\
2023-Oct-15 04:37:16.00 & $10.50 \pm 0.50$ & -- & 1.06 & 0.41 & AFA \\
2023-Oct-15 19:13:32.32 & $11.33 \pm 0.04$ & 12.89 & 1.07 & 0.42 & Unistellar \\
2023-Oct-15 21:35:33.00 & $9.23 \pm 0.97$ & 6.31 & 1.07 & 0.42 & AFA \\
2023-Oct-16 04:44:42.17 & $8.44 \pm 0.02$ & 12.18 & 1.07 & 0.42 & Unistellar \\
2023-Oct-16 07:29:55.00 & $10.09 \pm 0.11$ & 4.40 & 1.07 & 0.42 & AFA \\
2023-Oct-16 13:24:45.00 & $11.13$ & 3.26 & 1.07 & 0.42 & AFA \\
2023-Oct-16 13:29:03.00 & $9.65$ & 3.26 & 1.07 & 0.42 & AFA \\
2023-Oct-16 14:48:47.99 & $11.80 \pm 0.20$ & -- & 1.07 & 0.42 & AFA \\
2023-Oct-16 14:56:38.99 & $9.70 \pm 0.15$ & -- & 1.07 & 0.42 & AFA \\
2023-Oct-16 15:01:03.99 & $8.90 \pm 0.10$ & -- & 1.07 & 0.42 & AFA \\
2023-Oct-16 15:05:33.99 & $8.10 \pm 0.20$ & -- & 1.07 & 0.42 & AFA \\
2023-Oct-16 18:29:17.00 & $9.70$ & 0.98 & 1.07 & 0.42 & AFA \\
2023-Oct-16 18:49:19.00 & $10.25$ & 0.98 & 1.07 & 0.42 & AFA \\
2023-Oct-16 19:28:47.50 & $9.82 \pm 0.01$ & 1.79 & 1.07 & 0.42 & Unistellar \\
2023-Oct-16 19:29:08.99 & $9.59 \pm 0.40$ & 4.06 & 1.07 & 0.42 & AFA \\
2023-Oct-16 19:42:48.99 & $9.88 \pm 0.27$ & 4.17 & 1.07 & 0.42 & AFA \\
2023-Oct-16 20:09:57.00 & $8.88 \pm 0.28$ & 5.22 & 1.07 & 0.42 & AFA \\
2023-Oct-16 20:15:07.99 & $9.21 \pm 0.12$ & 5.24 & 1.07 & 0.42 & AFA \\
2023-Oct-17 03:09:05.19 & $16.45 \pm 0.16$ & 21.77 & 1.07 & 0.42 & Unistellar \\
2023-Oct-18 00:59:04.44 & $10.41 \pm 0.03$ & 19.13 & 1.07 & 0.42 & Unistellar \\
2023-Oct-18 03:54:36.99 & $9.41$ & 5.51 & 1.07 & 0.42 & AFA \\
2023-Oct-18 08:47:09.00 & $11.12$ & 0.45 & 1.07 & 0.42 & AFA \\
2023-Oct-18 10:33:26.99 & $12.25$ & 0.57 & 1.07 & 0.42 & AFA \\
2023-Oct-18 14:54:58.99 & $8.21$ & 0.37 & 1.07 & 0.42 & AFA \\
2023-Oct-18 15:06:05.00 & $9.20$ & 0.37 & 1.07 & 0.42 & AFA \\
2023-Oct-18 21:52:38.00 & $9.81 \pm 0.18$ & 6.11 & 1.07 & 0.42 & AFA \\
2023-Oct-19 13:33:40.99 & $9.56$ & 4.04 & 1.07 & 0.43 & AFA \\
2023-Oct-19 15:50:48.00 & $12.62$ & 0.49 & 1.07 & 0.43 & AFA \\
2023-Oct-19 16:04:50.00 & $12.17$ & 0.49 & 1.07 & 0.43 & AFA \\
2023-Oct-20 19:50:35.75 & $9.48 \pm 0.01$ & 11.53 & 1.07 & 0.43 & Unistellar \\
2023-Oct-21 17:22:15.16 & $9.26 \pm 0.01$ & 5.58 & 1.07 & 0.43 & Unistellar \\
2023-Oct-22 03:41:19.96 & $10.33 \pm 0.03$ & 21.12 & 1.07 & 0.44 & Unistellar \\
2023-Oct-22 05:38:28.99 & $10.73 \pm 0.14$ & 2.71 & 1.07 & 0.44 & AFA \\
2023-Oct-22 10:37:23.00 & $9.51 \pm 0.35$ & 5.45 & 1.07 & 0.44 & AFA \\
2023-Oct-22 18:59:43.89 & $10.01 \pm 0.11$ & 7.25 & 1.07 & 0.44 & Unistellar \\
2023-Oct-23 10:59:15.99 & $9.57$ & -- & 1.07 & 0.44 & AFA \\
2023-Oct-23 17:53:19.00 & $11.05 \pm 0.29$ & 1.38 & 1.08 & 0.44 & AFA \\
2023-Oct-24 09:08:50.00 & $9.14 \pm 0.21$ & 6.04 & 1.08 & 0.44 & AFA \\
2023-Oct-25 06:06:18.00 & $9.93 \pm 1.64$ & 2.86 & 1.08 & 0.45 & AFA \\
2023-Oct-29 03:42:20.88 & $12.54 \pm 0.02$ & 2.31 & 1.09 & 0.46 & Unistellar \\
2023-Nov-01 03:54:36.00 & $11.20 \pm 0.61$ & 1.10 & 1.10 & 0.47 & AFA \\
2023-Nov-01 17:49:39.38 & $13.51 \pm 0.02$ & 14.17 & 1.10 & 0.47 & Unistellar \\
2023-Nov-04 02:25:27.46 & $9.04 \pm 0.01$ & 22.51 & 1.11 & 0.48 & Unistellar \\
2023-Nov-07 18:50:43.82 & $10.42$ & 9.00 & 1.12 & 0.49 & Unistellar \\
2023-Nov-13 07:48:17.00 & $12.72 \pm 0.23$ & 1.10 & 1.15 & 0.51 & AFA \\
2023-Nov-13 16:41:23.75 & $10.32 \pm 0.04$ & 3.08 & 1.15 & 0.51 & Unistellar \\
2023-Nov-13 18:10:58.90 & $10.56 \pm 0.03$ & 20.40 & 1.15 & 0.51 & Unistellar \\
2023-Nov-14 04:13:59.75 & $9.07 \pm 0.01$ & 19.77 & 1.15 & 0.52 & Unistellar \\
2023-Nov-15 03:44:51.45 & $11.17 \pm 0.05$ & 10.74 & 1.16 & 0.52 & Unistellar \\
2023-Nov-15 08:16:52.00 & $12.35 \pm 1.14$ & 1.21 & 1.16 & 0.52 & AFA \\
2023-Nov-15 08:51:04.87 & $10.41 \pm 0.07$ & 6.84 & 1.16 & 0.52 & Unistellar \\
2023-Nov-16 09:28:10.65 & $9.39 \pm 0.02$ & 11.25 & 1.17 & 0.52 & Unistellar \\
2023-Nov-17 19:21:48.96 & $11.47 \pm 0.01$ & 0.07 & 1.17 & 0.53 & Unistellar \\
2023-Nov-19 05:02:58.85 & $11.64 \pm 0.01$ & 3.64 & 1.18 & 0.53 & Unistellar \\
2023-Nov-21 11:33:34.00 & $11.11 \pm 0.35$ & 3.21 & 1.20 & 0.54 & AFA \\
2023-Nov-22 04:40:23.92 & $10.65 \pm 0.01$ & 11.67 & 1.20 & 0.54 & Unistellar \\
2023-Nov-22 08:09:06.00 & $11.56 \pm 0.08$ & 2.93 & 1.20 & 0.54 & AFA \\
2023-Nov-23 04:05:20.66 & $9.40 \pm 0.03$ & 1.66 & 1.21 & 0.54 & Unistellar \\
2023-Nov-23 07:11:59.99 & $11.80 \pm 0.04$ & 2.95 & 1.21 & 0.54 & AFA \\
2023-Nov-23 16:29:45.00 & $11.65$ & 1.33 & 1.21 & 0.55 & AFA \\
2023-Nov-23 19:40:39.91 & $12.01 \pm 0.01$ & 12.94 & 1.21 & 0.55 & Unistellar \\
2023-Nov-25 04:07:36.51 & $12.56 \pm 0.30$ & 11.78 & 1.22 & 0.55 & Unistellar \\
2023-Nov-29 04:47:57.00 & $12.60 \pm 0.11$ & 2.97 & 1.25 & 0.56 & AFA \\
2023-Dec-02 17:29:43.43 & $12.35 \pm 0.01$ & 7.51 & 1.27 & 0.57 & Unistellar \\
2023-Dec-04 09:36:37.99 & $13.18$ & 0.63 & 1.28 & 0.58 & AFA \\
2023-Dec-05 09:51:07.00 & $12.17 \pm 0.04$ & 3.01 & 1.29 & 0.58 & AFA \\
2023-Dec-07 11:36:57.00 & $14.24$ & 13.90 & 1.31 & 0.58 & AFA \\
2023-Dec-13 01:49:59.80 & $13.52 \pm 0.02$ & 6.44 & 1.35 & 0.60 & Unistellar \\
2023-Dec-13 18:09:26.99 & $10.43 \pm 0.37$ & 4.56 & 1.35 & 0.60 & AFA \\
2023-Dec-13 18:37:21.99 & $10.96 \pm 0.09$ & 3.25 & 1.35 & 0.60 & AFA \\
2023-Dec-13 18:40:22.00 & $10.24 \pm 0.31$ & 4.01 & 1.35 & 0.60 & AFA \\
2023-Dec-19 22:06:31.00 & $10.13 \pm 0.03$ & 1.98 & 1.40 & 0.62 & AFA \\

\end{longtable}

\bibliography{main.bib}{}
\bibliographystyle{aasjournal}

\end{document}